\documentclass[fleqn,twoside]{article}

\usepackage{espcrc2}
\usepackage{epsf}

\title{
Multiple molecular dynamics time-scales in Hybrid Monte Carlo fermion 
simulations.}

\author{Mike Peardon\address[TCD]{The TrinLat Collaboration,\\
         School of Mathematics, Trinity College, Dublin 2, Ireland}
        \thanks{Talk presented by Mike Peardon}
        and 
        James Sexton\addressmark[TCD]}
       
\begin{document}

\begin{abstract}
A scheme for separating the high- and low-frequency molecular
dynamics modes in Hybrid Monte Carlo (HMC) simulations of gauge
theories with dynamical fermions is presented. The algorithm is tested
in the Schwinger model with Wilson fermions.
\vspace{1pc}
\end{abstract}

\maketitle

At this conference, much discussion focussed on the dominant
systematic error facing dynamical fermion simulations; the chiral
extrapolation.  It is proving to be a significant challenge to run Wilson
fermion simulations of QCD using the HMC algorithm at light quark masses
\cite{Irving:2002fx}.

The standard implementation of HMC introduces pseudofermion fields to mimic the
fermion action, and generates new proposals to a Metropolis test by integrating
the molecular-dynamics (MD) equations of motion for a Hamiltonian in a
fictitious simulation time.  The maximum step-size for a useful Metropolis
acceptance rate is set by a characteristic time-scale of the action used in the
Hamiltonian \cite{Joo:2000dh}.  Recent studies find that as the fermion is made
lighter, the force induced by the pseudofermions generates increasingly violent
high-frequency fluctuations.  This effect is called ``ultra-violet
slowing-down'' in Ref.  \cite{Borici:2002zu}. In this report, we describe a
modification to HMC that attempts to address this problem.  A number
of interesting algorithms have been devised recently that reduce the influence 
of the UV fermion modes on either the Monte Carlo update scheme
or in the lattice discretisation
\cite{Borici:2002zu,Duncan:1998gq,Hasenfratz:2002jn,Orginos:1999cr,Alexandrou:1999ii}.

\section{MULTIPLE MD TIME-SCALES}

A scheme for integrating the equations of motion for the
MD phase of the HMC algorithm by introducing different
time-scales for different segments of the action was introduced in Ref.
\cite{Sexton:1992nu}.

The leap-frog integrator is constructed from the two simple time-evolution
operators generated by the the kinetic and potential energy terms. Their effect 
on the system coordinates, $\{p,q\}$ are
\begin{eqnarray}
V_T(\Delta\tau): \{p,q\} &\longrightarrow& \{p,q + \Delta \tau \;p\} \nonumber\\
V_S(\Delta\tau): \{p,q\} &\longrightarrow& \{p - \Delta \tau\;\partial S,q\}
\end{eqnarray}
The simplest reversible leap-frog integrator is then 
\begin{equation}
  V(\Delta \tau) =
       V_S(\frac{\Delta \tau}{2})
       V_T(\Delta \tau)
       V_S(\frac{\Delta \tau}{2}).
\end{equation}
If the action (and thus the Hamiltonian) is split into two parts, 
\begin{eqnarray}
   {\cal H} = \underbrace{T(p) + S_1(q)}_{{\cal H}_1}
            + \underbrace{S_2(q)}_{{\cal H}_2}, 
\end{eqnarray}
then the two leap-frog integrators for these two pieces can be
constructed as 
\begin{eqnarray}
V_1(\Delta \tau) &=&
       V_{S_1}(\frac{\Delta \tau}{2})
       \;V_T(\Delta \tau)\;
       V_{S_1}(\frac{\Delta \tau}{2}) \nonumber \\
V_2(\Delta \tau) &=& V_{S_2}(\Delta \tau)
\end{eqnarray}
and a reversible integrator for the full Hamiltonian can be constructed by 
combining the two schemes:
\begin{equation}
  V(\Delta \tau) =
       V_2(\frac{\Delta \tau}{2})
       \;\left[V_1(\frac{\Delta\tau}{m}) \right]^m\;
       V_2(\frac{\Delta \tau}{2}),
\end{equation}
with $m\in Z$. 
This compound integrator effectively introduces two evolution time-scales, 
$\Delta\tau$ and $\Delta\tau/m$.

We suggest that the multiple time-scale scheme is only helpful when two 
criteria are fulfilled simultaneously:
\begin{enumerate}
  \item{the force term generated by $S_1$ is cheap to compute compared to that
of $S_2$ and}
  \item{the split captures the high-frequency modes of the system in $S_1$ and 
the low-frequency modes in $S_2$.}
\end{enumerate}
The popular implementation of the scheme in dynamical fermion simulations with
pseudofermions is to split the Hamiltonian into the Yang-Mills term and the
pseudofermion action:
\begin{eqnarray}
  S_1 &=& S_G, \nonumber \\
  S_2 &=& \phi^* [M^\dagger M]^{-1} \phi
\end{eqnarray}
Unfortunately for light fermions, the highest frequency fluctuations are in the
pseudofermion action, which also has the more computationally expensive force
term, thus the criteria are not met.  The key to using the method then is to
implement a low-computational cost scheme that separates the high- and
low-frequency fermion molecular-dynamics.

\section{POLYNOMIAL FILTERING}

\begin{figure}[t]
 \setlength{\epsfxsize}{7.4cm}
 \vspace{0.5ex}\epsfbox{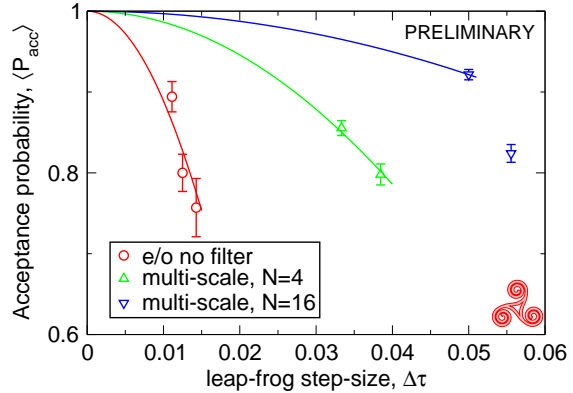}
 \caption{The HMC acceptance rate as a function of MD step-size for the
standard leap-frog method, and for the new algorithm with $N=4$ and $16$
  \label{fig:pacc}}
\end{figure}

A very low-order polynomial approximation offers a cheap means of mimicking
most of the short-distance physics of the fermion interations. Recent interest
in polynomial approximations to a matrix inverse was inspired by the
multi-boson algorithm of Ref. \cite{Luscher:1993xx}. This led to the
development of the Polynomial HMC (PHMC) algorithm \cite{Frezzotti:1997ym}.
Following this idea, we write an exact representation of the two-flavour
probability measure
\begin{equation}
\det M^\dagger M  =  \int\!\!
      {\cal D}\phi
      {\cal D}\phi^*
      {\cal D}\chi
      {\cal D}\chi^*
           e^{\left\{-S_\phi-S_\chi\right\}}
\end{equation}
with 
\begin{equation}
S_\phi = \left|[M {\cal P}(M)]^{-1} \phi \right|^2
   \mbox{ and } 
S_\chi = \left| {\cal P}(M) \chi \right|^2
  \label{eqn:action}
\end{equation}

The fields $\phi$ are modified pseudofermions and we term the new fields
$\chi$, ``guide'' bosons.  The algorithm exactly recovers the probability
measure for the two-flavour theory for any choice of polynomial.  This scheme
is similar to the split-pseudofermion method of Ref. \cite{Hasenbusch:2001ne}
with one distinction; using polynomials allows us to modify the fermion modes
easily and cheaply.  The success of the algorithm hinges on the empirical
observation that the fermion modes that induce the high-frequency fluctuations
are localised and can be handled by very low-order polynomials. This means the
force term for the guide bosons, generated by $S_\chi$ can be computed cheaply
and a multiple time-scale integrator can be built that simultaneously fulfills
the two criteria discussed earlier. The time-scale split is then 
\begin{equation}
  S = 
    \underbrace{S_G + S_\chi}_{UV,\;\; \partial S\mbox{ cheap}} + 
    \underbrace{S_\phi}      _{IR,\;\; \partial S\mbox{ costly.}}
\end{equation}

Note that in general, $p$ time-scales can be introduced straightforwardly 
by first adding $p-1$ sets of guide fields, then 
constructing a heirarchy of leapfrog integrators, with $m_i$ leap-frog steps at 
the $i^{\rm th}$ level. This introduces $p$ time-scales, $\Delta\tau > 
\frac{\Delta\tau}{m_{p-1}} \dots 
\frac{\Delta\tau}{m_{p-1} \dots m_{1}}$.

\section{TESTING THE ALGORITHM}

The performance of the method is under investigation in simulations of the 
two-flavour massive Schwinger model ($2d$ QED). HMC runs presented here are 
performed on $64\times64$ lattices, with $\beta=4.0, \kappa = 0.2618$.
Non-hermitian Chebyshev polynomial approximations are used \cite{Borici:1995am}.
Measurements of the dependence of the acceptance probability on the polynomial
degree and tuning the MD parameters, $\Delta\tau$ and $m$ are being made. 

Fig.~\ref{fig:pacc} shows the acceptance rate for the standard HMC algorithm
(with even-odd preconditioned pseudofermions) along with the 
polynomial-filtered method. Two different polynomials were used; $N=4$ and
$N=16$. The solid lines are fits to the expected behaviour of HMC 
as $\Delta\tau \rightarrow 0$, namely
\begin{equation}
  \langle P_{\rm acc}\rangle = \mbox{erfc}\left\{(\Delta\tau/\tau_0)^2\right\},
\end{equation}
where the value of $\tau_0$ is determined by the best fit. $\tau_0$ is then a 
characteristic time-scale for the modes encapsulated in the pseudofermionic
part of the action, $S_\phi$ in Eqn. \ref{eqn:action}. Since evaluation of the
force term $\partial S_\phi$ dominates in this parameter range, $\tau_0$ is a
good indicator of algorithm performance.

\section{TUNING THE ALGORITHM}

\begin{figure}[t]
 \setlength{\epsfxsize}{7.4cm}
 \vspace{0.5ex}\epsfbox{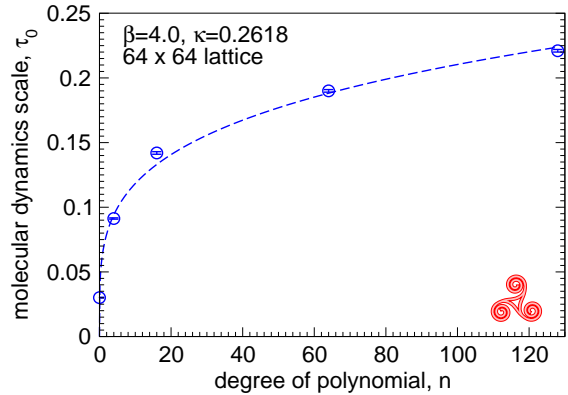}
 \caption{The characteristic time-scale $\tau_0$ vs polynomial degree, $N$. The 
     dashed line guides the eye. \label{fig:tau0}}
\end{figure}

The algorithm has a number of free parameters, allowing a good deal of scope to
optimise the performance. Fig.~\ref{fig:tau0} shows the characteristic
time-scale for pseudofermion integration, $\tau_0$ as a function of Chebyshev
polynomial degree. $\tau_0$ rises very rapidly initially as the degree of the
filter algorithm is increased, and $\tau_0$ for $N=16$ is about four times
larger than for the standard HMC algorithm. Inverting the pseudofermion matrix
requires roughly the same computational cost, and thus the algorithm is four
times more efficient than HMC (assuming autocorrelations are the same). 

For low values of $N$ (the polynomial degree) the computational bottleneck is
solver performance while for larger values, the evaluation of the force term
arising from interactions between the gauge fields and the guide bosons begins
to dominate. Effective and simple strategies for tuning the algorithm are still
under investigation. We are also investigating alternative choices of
polynomial filters beyond Chebyshev approximation. 

MP is grateful to Enterprise-Ireland for support under grant SC/01/306.

\end{document}